\documentclass[sigconf,nonacm]{acmart}

\usepackage{amsmath,amsthm}
\usepackage{multirow,colortbl}
\usepackage{float}
\floatstyle{ruled}
\newfloat{algorithm}{tbp}{loa}
\floatname{algorithm}{Algorithm}
\floatstyle{plain}
\usepackage{tikz}
\usetikzlibrary{arrows.meta,positioning,calc}

\DeclareMathOperator{\sg}{sg}

\DeclareMathOperator{\RelayTop}{Top}

\DeclareMathOperator{\norm}{norm}
\newcommand{\ind}[1]{\mathbf{1}\{#1\}}

\definecolor{OCBlue}{HTML}{356A9A}
\definecolor{OCTeal}{HTML}{187F77}
\definecolor{OCOrange}{HTML}{B45C22}
\definecolor{OCPurple}{HTML}{7C4B94}
\definecolor{OCTableBestBaseline}{HTML}{FFF0D6}
\definecolor{OCTableOurs}{HTML}{EAF2F8}
\definecolor{OCTableHeader}{HTML}{F2F5F8}
\definecolor{OCTableCategoryText}{HTML}{52616E}
\newsavebox{\ocmaintablemeasure}
\newcommand{\OCIgnoreTableColor}[1]{}
\newcommand{\OCMainTableFit}[3]{%
  \begingroup
  \setlength{\tabcolsep}{0pt}%
  \sbox{\ocmaintablemeasure}{%
    \let\rowcolor\OCIgnoreTableColor
    \let\cellcolor\OCIgnoreTableColor
    \begin{tabular}{#2}#3\end{tabular}}%
  \setlength{\tabcolsep}{\dimexpr
    (\linewidth-\wd\ocmaintablemeasure)/\numexpr 2*#1\relax\relax}%
  \begin{tabular}{#2}#3\end{tabular}%
  \endgroup
}
\newcommand{\OCMeanSD}[2]{\shortstack[c]{\textbf{#1}\\[-2pt]{\scriptsize$\pm$\,#2}}}
\newlength{\octermwidth}
\newlength{\oclabelwidth}
\newcommand{\octerm}[4][\displaystyle]{%
  \begingroup\setlength{\fboxsep}{1.2pt}%
  \settowidth{\octermwidth}{\colorbox{#2!10}{$#1 #3$}}%
  \settowidth{\oclabelwidth}{$\scriptstyle\text{\normalfont\bfseries #4}$}%
  \ifdim\oclabelwidth>\octermwidth
    \setlength{\octermwidth}{\oclabelwidth}%
  \fi
  \underbrace{\makebox[\octermwidth][c]{%
    \colorbox{#2!10}{$#1 #3$}}}%
  _{\textcolor{#2}{\text{\normalfont\bfseries #4}}}%
  \endgroup
}
\usepackage{tabularx,colortbl,float}
\providecommand{\sg}{\operatorname{sg}}
\providecommand{\norm}{\operatorname{norm}}
\providecommand{\RelayTop}{\operatorname{Top}}
\providecommand{\ind}[1]{\mathbf 1\{#1\}}
\definecolor{RelayBlue}{HTML}{1F4F7A}
\definecolor{RelayTeal}{HTML}{0F655D}
\definecolor{RelayOrange}{HTML}{914515}
\newcommand{\rterm}[3]{%
  \underbrace{\colorbox{#1!10}{$\displaystyle #2$}}_{\text{\scriptsize\textcolor{#1}{#3}}}%
}
\makeatletter
\@ifundefined{algorithm}{%
  \floatstyle{ruled}%
  \newfloat{algorithm}{tbp}{loa}%
  \floatname{algorithm}{Algorithm}%
  \floatstyle{plain}%
}{}
\makeatother

\setcopyright{none}
\begin{document}

\title[RELAY]{RELAY: One Query, Two Roles for Sequential Recommendation}
\author{Zesheng Li}
\affiliation{%
  \institution{University of Chinese Academy of Sciences}
  \city{Beijing}
  \country{China}}
\email{lizesheng26@mails.ucas.ac.cn}
\author{Chengchang Pan}
\affiliation{%
  \institution{University of Chinese Academy of Sciences}
  \city{Beijing}
  \country{China}}
\email{166353314@qq.com}
\author{Honggang Qi}
\authornote{Corresponding author.}
\affiliation{%
  \institution{University of Chinese Academy of Sciences}
  \city{Beijing}
  \country{China}}
\email{hgqi@ucas.ac.cn}
\renewcommand{\shortauthors}{Li et al.}

\begin{abstract}
Sequential recommendation must combine semantic preferences with behavioral associations to deliver accurate rankings at low inference cost. Sharing history selection connects preference representation directly to the behavioral evidence used in candidate scoring. We present RELAY, a compact sequential recommender built around \emph{one query, two roles}: semantic readout and behavioral-memory routing. A single learned query forms a semantic preference representation and reuses its attention to route direct-history and transition evidence into candidate scores. Candidate-conditioned gates calibrate the two memory channels for each candidate. RELAY trains both roles with a single catalog likelihood and produces the final ranking in one forward pass. Full-catalog experiments on six datasets show that RELAY achieves the highest Recall@5 and NDCG@5 among the compared methods on every dataset, improving over the previous best model by $14.22\%$ and $17.88\%$ on average. With $2.044$ million predictor parameters, RELAY reduces predictor size by $65.12\%$ and achieves a $1.75$--$1.79\times$ inference speedup over the previous best model.
\end{abstract}


\maketitle
\hypersetup{pdfauthor={Zesheng Li, Chengchang Pan, Honggang Qi}}

\section{Introduction}
\label{sec:introduction}

Sequential recommendation on the Web must turn a user's interaction history into an accurate recommendation within a tight inference budget. That history carries two complementary forms of evidence. Shared item semantics reveal broader preferences, while observed behavioral associations identify specific choices that tend to follow past interactions~\cite{hou2023vqrec,rajput2023tiger,rendle2010fpmc,zhu2024mqsated}. Neural sequence models summarize historical preferences for next-item prediction~\cite{hidasi2016gru4rec,kang2018sasrec,sun2019bert4rec}. The modeling goal is to make semantic preferences and explicit behavioral evidence cooperate in one compact scoring path.

\begin{figure}[t]
\centering
\includegraphics[width=\linewidth]{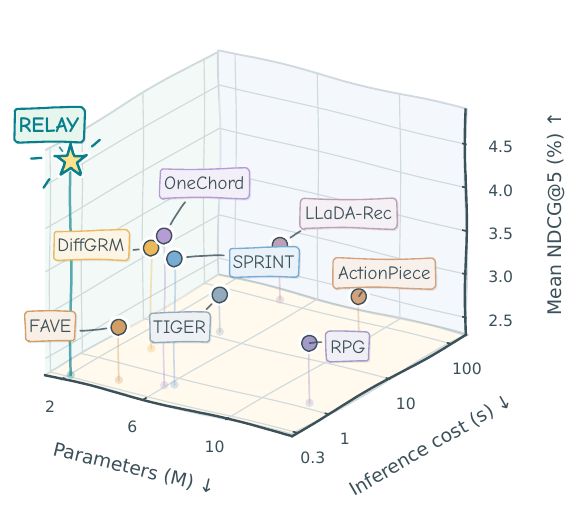}
\caption{Accuracy, predictor size, and inference cost, averaged equally over six datasets. Cost uses a logarithmic axis. RELAY combines the highest mean NDCG@5 with the smallest predictor and lowest inference cost among the plotted models.}
\Description{A hand-drawn-style three-dimensional scatter plot compares nine recommenders on mean NDCG at five, mean predictor parameters in millions, and mean full-test inference cost in seconds. The cost axis is logarithmic. A gold star with a teal outline marks RELAY at higher accuracy, fewer parameters, and lower cost than FAVE and the seven plotted SID-based baselines. Every point has a rounded label with its method name.}
\label{fig:relay-teaser}
\end{figure}

Semantic readout and behavioral-memory routing share a central decision: \emph{which historical interactions matter for the current recommendation}. Semantic readout aggregates relevant interactions into a preference representation; memory routing uses those interactions to supply associations for candidate scoring. For example, in a history mixing LEGO themes, attention to a NINJAGO interaction expresses a theme preference and routes its observed transition evidence to specific candidate sets. Sharing their history selection lets the final ranking objective jointly determine what represents a user's preferences and which behavioral evidence supports each candidate. This motivates a simple principle: \textbf{select the history once, use its evidence twice}.

We present \textbf{RELAY}\footnote{Code: \url{https://github.com/LIZESHENG13/RELAY}.}, a sequential recommender built around \textbf{one query, two roles: semantic readout and memory routing}. A single learned query reads the history to form a semantic preference representation and reuses the same history attention to route direct-history and transition evidence into candidate scores. Semantic matching scores the complete catalog, while candidate-conditioned gates calibrate the memory evidence for each candidate. A common catalog likelihood trains both roles through the final ranking scores, and recommendation takes one forward pass.

Figure~\ref{fig:relay-teaser} summarizes RELAY's joint gains in ranking quality, predictor size, and inference time. Our contributions are:
\begin{list}{\small\textbullet}{%
  \setlength{\leftmargin}{0.85em}%
  \setlength{\labelwidth}{0.4em}%
  \setlength{\labelsep}{0.45em}%
  \setlength{\itemindent}{0pt}%
  \setlength{\listparindent}{0pt}%
  \renewcommand{\makelabel}[1]{\makebox[\labelwidth][l]{#1}}%
  \setlength{\topsep}{2pt}%
  \setlength{\partopsep}{0pt}%
  \setlength{\itemsep}{2pt}%
  \setlength{\parsep}{0pt}}%
\item We introduce \textbf{one query, two roles} for semantic readout and behavioral-memory routing. Shared history selection links preference representation directly to candidate-level ranking evidence.
\item We train both roles with one catalog likelihood and derive how the final ranking loss supervises shared attention through semantic readout and explicit memory routing.
\item RELAY achieves the highest Recall@5 and NDCG@5 across the evaluated domains, improving over the previous best model by \textbf{14.22\%} and \textbf{17.88\%} on average, with \textbf{65.12\%} fewer predictor parameters and a \textbf{1.75--1.79$\times$} inference speedup.
\end{list}

\section{Related Work}
\label{sec:related_work}

\paragraph{\textbf{Sequential preference modeling.}}
Recurrent and attention-based recommenders learn user preferences from interaction sequences~\cite{hidasi2016gru4rec,kang2018sasrec,sun2019bert4rec}. Their history encoders capture dependencies among past actions and provide representations for next-item prediction. Multi-interest models use dynamic routing or attention to extract several preference vectors, supporting diverse interests and controllable retrieval~\cite{li2019mind,cen2020comirec}. These studies establish history selection and aggregation as central modeling choices. RELAY assigns a single query two coupled responsibilities: its attention aggregates the semantic representation and allocates explicit memory evidence to catalog candidates. The shared selection connects representation learning directly to the local evidence used in ranking.

\paragraph{\textbf{Semantic representations and direct ranking.}}
Content-aware sequence learning and vector-quantized item representations share information across items and recommendation domains~\cite{hou2022unisrec,hou2023vqrec}. Generative retrieval represents items with short semantic identifiers and predicts their tokens from user histories~\cite{rajput2023tiger}. Tokenization research further incorporates collaborative semantics, code diversity, sequence context, and joint optimization of the index and recommender, including differentiable semantic-ID learning~\cite{zheng2024lcrec,wang2024letter,hou2025actionpiece,liu2025etegrec,fu2026diger}. Semantic codes also support direct catalog scoring: DiGR composes SID representations and optimizes item-level matching~\cite{zhao2026digr}. RELAY uses SID representations within a unified catalog scorer. Its focus is the shared attention that links a semantic preference vector to bounded, explicit behavioral evidence.

\paragraph{\textbf{Transition and history memory.}}
Personalized Markov models combine user preferences with item transitions~\cite{rendle2010fpmc}; graph-based methods encode behavioral relations in item and session representations~\cite{wu2019srgnn,li2026bridge}. Repeat-aware recommendation makes historical items explicitly available to a recommendation mechanism~\cite{ren2019repeatnet}, and transition-aware distillation transfers global transition patterns into item embeddings~\cite{zhu2024mqsated}. RELAY retains a bounded transition memory at scoring time and routes its contributions with the attention already used for semantic readout. It also addresses a specific dependence introduced by aggregated transition counts: each training record contributes to the memory used to score that record. Target-statistic research studies related dependence between labels and derived features~\cite{prokhorenkova2018catboost}. RELAY removes the current record's one-count contribution and exactly restores its top-neighbor set before computing the training loss.

\paragraph{\textbf{Efficient recommendation and one-pass inference.}}
Parallel SID prediction reduces sequential decoding, while structured search recovers legal catalog identifiers~\cite{hou2025rpg,wang2026onegr}. Discrete diffusion provides another route to parallel SID modeling~\cite{shi2025lladarec,liu2026diffgrm}, and single-step prediction directly estimates token probabilities for efficient catalog ranking~\cite{cai2026sprint}. Distillation transfers ranking information or teacher predictions into compact recommenders~\cite{tang2018rd,guo2026sidmlp}; OneChord learns and distills a fused catalog distribution~\cite{li2026onechord}. RELAY brings these ideas into one full-catalog scorer: semantic matching assigns a score to every SID, while query-routed memory adjusts candidates reached by direct history and bounded transition neighborhoods. One catalog likelihood trains both terms, connecting preference estimation to candidate-level behavioral evidence.

\section{Method}
\label{sec:method}

\begin{figure*}[t]
\centering
\includegraphics[width=0.86\textwidth]{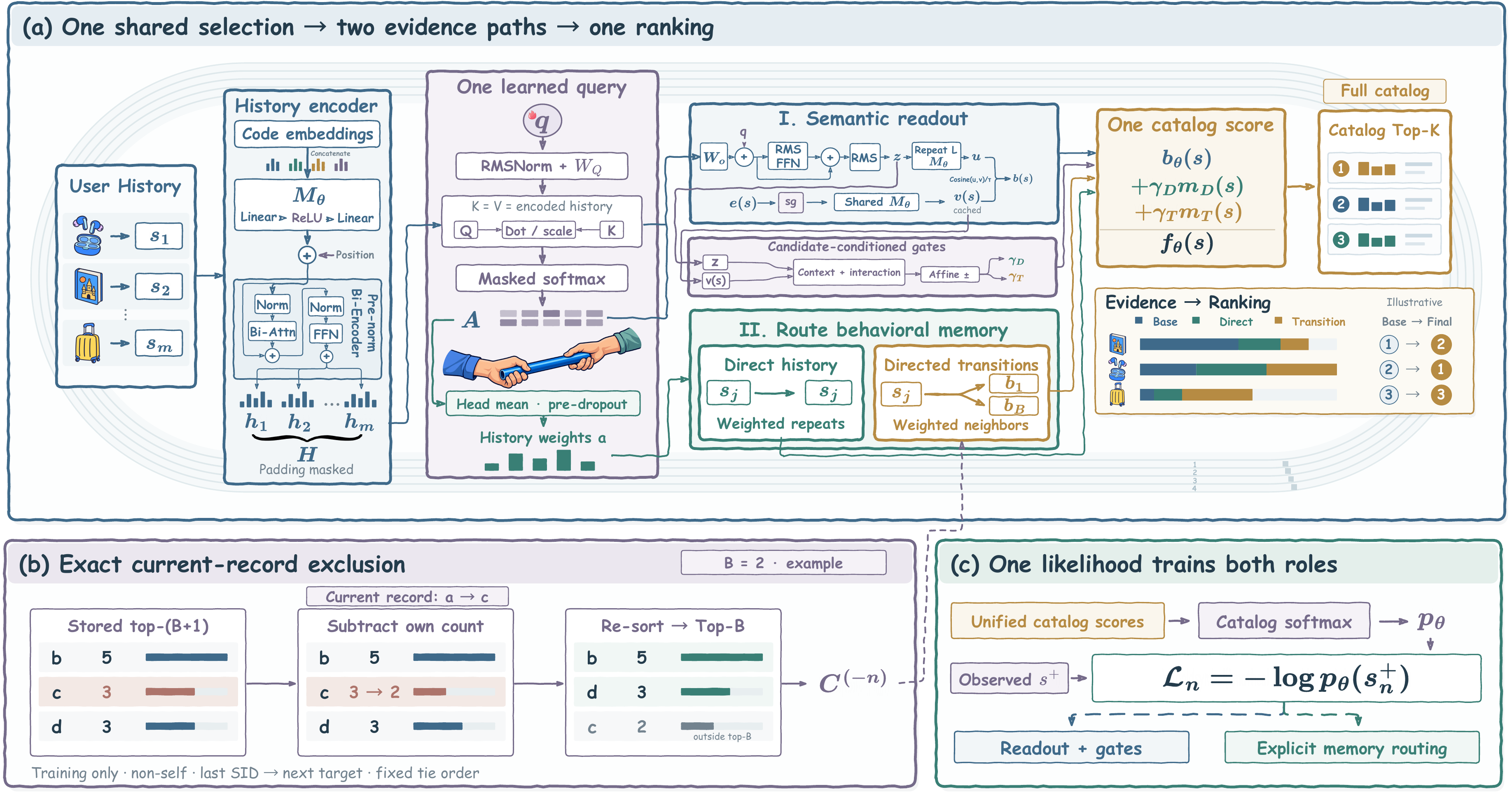}
\caption{Overview of RELAY. (a) One learned query selects the encoded history for two roles: semantic readout and behavioral-memory routing. Their evidence forms one catalog score for Top-$K$ recommendation. (b) Training removes the current record's transition contribution and recovers the adjusted top-$B$ neighbors from a top-$(B+1)$ buffer. (c) A single catalog likelihood trains the readout, context-dependent memory strengths, and shared routing attention.}
\Description{RELAY framework with three panels. A single query reads the encoded history and shares its attention with direct-history and transition memory. Semantic matching and memory corrections combine into full-catalog ranking scores. A training panel illustrates exact neighbor replacement after removing the current record's transition count. A common catalog likelihood supervises both query roles.}
\label{fig:relay-framework}
\end{figure*}

RELAY follows one principle: \textbf{select the history once, use its evidence twice}. A single query reads a semantic preference representation and routes behavioral memory with the same attention. The resulting evidence enters one catalog score, so the ranking objective jointly learns what to read and where to apply it. Figure~\ref{fig:relay-framework} presents the framework.

\subsection{Semantic Catalog and History Representation}
\label{sec:relay-catalog}

For an interaction history $H=(i_1,\ldots,i_m)$ and next item $i^+$, a fixed semantic index assigns an $L$-code identifier:
\begin{equation}
\begin{aligned}
\sigma(i)&=(s_1,\ldots,s_L)\in\textstyle\prod_{\ell=1}^{L}\mathcal V_\ell,\\
\mathcal S&=\{\sigma(i):i\in\mathcal I\},\qquad s^+=\sigma(i^+).
\end{aligned}
\label{eq:relay-catalog}
\end{equation}
The distinct tuples in $\mathcal S$ define the candidate catalog; items assigned the same tuple share a candidate. Write $s_j=\sigma(i_j)$. With code embeddings $E_\ell$ and a shared two-layer projection $M_\theta:\mathbb R^{Ld}\!\to\mathbb R^d$, the history encoder computes
\begin{equation}
\begin{aligned}
e_\theta(s)&=[E_1(s_1);\ldots;E_L(s_L)],\\
x_j&=M_\theta(e_\theta(s_j))+p_j,\\
(h_1,\ldots,h_m)&=\operatorname{Enc}_\theta(x_1,\ldots,x_m).
\end{aligned}
\label{eq:relay-history}
\end{equation}
Here $p_j$ denotes the learned embedding of the interaction's position in the padded history. The bidirectional encoder contextualizes each interaction; padding is masked throughout. $M_\theta$ has a ReLU hidden layer and is shared by history encoding, semantic readout, and candidate representation.

\subsection{One Query, Two Roles}
\label{sec:relay-query}

\paragraph{\textbf{One shared history selection.}}
Let $q\in\mathbb R^d$ be a learned query, $J$ the number of attention heads, and $d_h=d/J$. For head $h$, let $q_h$ be the corresponding slice of $W_Q\operatorname{RMSNorm}(q)$ and $h_{jh}$ the head slice of $h_j$. The encoded history supplies both keys and values directly. The query defines
\begin{equation}
\begin{aligned}
\eta_{hj}&=q_h^\top h_{jh}/\sqrt{d_h},\\
A_{hj}&=\frac{\exp\eta_{hj}}{\sum_{k=1}^{m}\exp\eta_{hk}},\\
a_j&=\rterm{RelayTeal}{\frac{1}{J}\sum_{h=1}^{J}A_{hj}}{Shared routing weight}.
\end{aligned}
\label{eq:relay-attention}
\end{equation}
Because each head distribution is normalized, $a_j\geq0$ and $\sum_j a_j=1$. Contextual keys give each history its own allocation, so repeated occurrences of one SID can receive different weights. These shared weights drive both semantic readout and memory routing; longer histories distribute the same total mass across more positions.

\paragraph{\textbf{Role I: semantic readout.}}
The head contexts form one query state through a residual cross-attention block and a feed-forward block:
\begin{equation}
\begin{aligned}
c_h&=\sum_{j=1}^{m}A_{hj}h_{jh},\\
z'&=q+W_O[c_1;\ldots;c_J]+b_O,\\
z&=\operatorname{RMSNorm}\!\left(z'
+\operatorname{FFN}(\operatorname{RMSNorm}(z'))\right).
\end{aligned}
\label{eq:relay-readout}
\end{equation}
Standard training dropout is omitted from the notation; memory routing always uses the normalized attention before dropout. The readout operates on one query state. Its normalized semantic representation and the candidate representation are
\begin{equation}
\begin{aligned}
u_\theta(H)&=\norm\!\big(M_\theta(
\underbrace{[z;\ldots;z]}_{L\ \text{copies}})\big),\\
v_\theta(s)&=\norm\!\left(M_\theta(\sg[e_\theta(s)])\right),\\
b_\theta(s\mid H)&=\rterm{RelayBlue}{u_\theta(H)^\top v_\theta(s)}{Semantic compatibility}/\tau.
\end{aligned}
\label{eq:relay-semantic-score}
\end{equation}
Here $\norm(x)=x/\|x\|_2$ and $\tau>0$. The stop-gradient $\sg$ applies only to candidate-side code embeddings; the shared projection remains trainable.
Semantic matching therefore assigns a score to every catalog SID, including candidates without a direct history or transition match.

\paragraph{\textbf{Role II: memory routing.}}
The same $a_j$ weights the behavioral evidence contributed by each interaction. An interaction can support its own SID through direct history and associated SIDs through transitions, using the same allocation that forms the semantic readout.

\subsection{Attention-Routed Behavioral Memory}
\label{sec:relay-memory}

\paragraph{\textbf{Directed transition memory.}}
Let $\mathcal D_{\mathrm{tr}}$ contain $N$ training records $(H_n,s_n^+)$. Write $s_{n,\mathrm{last}}$ for the final SID in $H_n$. Off-diagonal transition counts are
\begin{equation}
C(a,b)=\sum_{n=1}^{N}
\ind{s_{n,\mathrm{last}}=a,\ s_n^+=b,\ a\ne b}.
\label{eq:relay-counts}
\end{equation}
For each source $a$, $\mathcal N_B^C(a)$ retains its $B$ largest positive counts, resolving ties by a fixed catalog order. The memory weight is
\begin{equation}
w_C(a,b)=
\begin{cases}
\log(1+C(a,b)),&b\in\mathcal N_B^C(a),\\
0,&\text{otherwise}.
\end{cases}
\label{eq:relay-transition-weight}
\end{equation}
The logarithm compresses count magnitude, while the neighbor budget bounds the evidence retrieved from any history interaction. Self transitions are represented through direct history evidence.

\paragraph{\textbf{Two memory channels, one routing decision.}}
Direct history evidence identifies previously observed SIDs; transition evidence connects historical SIDs to associated choices:
\begin{equation}
\begin{aligned}
d_j(s)&=\ind{s=s_j},&
t_j(s;C)&=w_C(s_j,s),\\
m_{\mathrm D}(s\mid H)&=\sum_{j=1}^{m}a_jd_j(s),&
m_{\mathrm T}(s\mid H;C)&=\sum_{j=1}^{m}a_jt_j(s;C).
\end{aligned}
\label{eq:relay-memory-channels}
\end{equation}
A repeated SID receives the sum of attention weights at all matching history positions. Transition evidence also accumulates when several historical SIDs connect to the same candidate; each source contributes its attention weight times the corresponding log-count. Thus, two history SIDs with a shared outgoing neighbor jointly support that candidate, while contextual attention can weight their contributions differently. Both channels use the same normalized history allocation. The query state and each candidate vector then determine their separate strengths through candidate-conditioned gates. Let $w_c$ and $b_c$ be row $c$ of $W_\lambda$ and entry $c$ of $b_\lambda$, where $c\in\{\mathrm D,\mathrm T\}$. The gates are
\begin{equation}
\begin{aligned}
x_c(s,z)&=\rterm{RelayTeal}{w_c^\top z+b_c}{History term}\\
&\quad+\rterm{RelayOrange}{\sqrt d\,(w_c\odot z)^\top v_\theta(s)}{Candidate interaction},\\
\gamma_c(s,z)&=\tfrac{1}{2}x_c(s,z)+\log 2.
\end{aligned}
\label{eq:relay-strengths}
\end{equation}
The shared attention selects the evidence; these gates calibrate each channel to the current candidate. The unified score is
\begin{equation}
\begin{aligned}
f_\theta(s\mid H;C)
&=\rterm{RelayBlue}{b_\theta(s\mid H)}{Semantic matching}\\
&\quad+\rterm{RelayTeal}{\gamma_{\mathrm D}(s,z)m_{\mathrm D}(s\mid H)}{Direct history}\\
&\quad+\rterm{RelayOrange}{\gamma_{\mathrm T}(s,z)m_{\mathrm T}(s\mid H;C)}{Directed transitions}.
\end{aligned}
\label{eq:relay-unified-score}
\end{equation}
The semantic term scores every catalog candidate; direct-history and transition terms contribute where the corresponding evidence is present. Their candidate-conditioned strengths are learned together with the shared query under the catalog likelihood.

\subsection{Record-Excluded Catalog Learning}
\label{sec:relay-learning}

\paragraph{\textbf{Learning from other records' transitions.}}
Each non-self training transition contributes one count to Eq.~\eqref{eq:relay-counts}. To train its predictor using evidence from the remaining records, define
\begin{equation}
\begin{aligned}
C^{(-n)}(a,b)
&=\rterm{RelayBlue}{C(a,b)}{All records}\\
&\quad-\rterm{RelayOrange}{\ind{a=s_{n,\mathrm{last}},\ b=s_n^+,\ a\ne b}}{Record~$n$}.
\end{aligned}
\label{eq:relay-record-exclusion}
\end{equation}
Training selects neighbors from $C^{(-n)}$ after the decrement. The adjusted row applies to every occurrence of its source SID in $H_n$; direct-history evidence still comes from the observed history.

\begin{proposition}[Exact neighbor replacement]
\label{prop:relay-exclusion}
Suppose one count in a source row decreases by one, and neighbor ordering uses descending count followed by a fixed tie order. The updated top-$B$ positive-count neighbors can be recovered exactly from the original top-$(B+1)$ neighbors and their adjusted counts.
\end{proposition}

A single decrement can move only one entry down, so its first possible replacement is already ranked $B+1$. Retaining the top $B+1$ during training therefore suffices: decrement, rerank, and keep the top $B$, without scanning the catalog row. The proof is in Appendix~\ref{app:relay-exclusion}.

\paragraph{\textbf{One likelihood for the complete ranking.}}
Define a catalog distribution from the unified score:
\begin{equation}
p_\theta(s\mid H;C)=
\frac{\exp f_\theta(s\mid H;C)}
{\rterm{RelayBlue}{\sum_{v\in\mathcal S}\exp f_\theta(v\mid H;C)}{Full-catalog normalizer}}.
\label{eq:relay-distribution}
\end{equation}
All trainable components are optimized by the next-target likelihood
\begin{equation}
\mathcal L(\theta)=-\frac{1}{N}\sum_{n=1}^{N}
\log p_\theta(s_n^+\mid H_n;C^{(-n)}).
\label{eq:relay-objective}
\end{equation}
The full-catalog likelihood jointly trains query attention, semantic representations, and both candidate-conditioned gates; transition counts remain fixed statistics. At recommendation time, the full-data $C$ supplies memory and $f_\theta$ ranks the catalog.

\paragraph{\textbf{How one query learns both roles.}}
For one training record, write $\ell=-\log p_\theta(s^+\mid H;C^{(-n)})$ and define
\begin{equation}
\begin{aligned}
\delta_s&=p_\theta(s\mid H;C^{(-n)})-\ind{s=s^+},\\
R_j(s;z)&=\gamma_{\mathrm D}(s,z)d_j(s)+\gamma_{\mathrm T}(s,z)t_j(s;C^{(-n)}),\\
\varepsilon_j&=\sum_{s\in\mathcal S}\delta_s R_j(s;z).
\end{aligned}
\label{eq:relay-routing-residual}
\end{equation}
$\varepsilon_j$ measures the ranking residual associated with the evidence routed by position $j$. Let $\xi_{hj}$ denote the derivative through the readout state $z$, including its effect on semantic scores and candidate-conditioned gates. Put $\bar\xi_h=\sum_k A_{hk}\xi_{hk}$ and $\bar\varepsilon_h=\sum_k A_{hk}\varepsilon_k$.

\begin{proposition}[Coupled supervision of shared attention]
\label{prop:relay-shared-gradient}
For valid history positions, the attention-logit gradient is
\begin{equation}
\begin{aligned}
\frac{\partial\ell}{\partial\eta_{hj}}
&=\rterm{RelayBlue}{A_{hj}(\xi_{hj}-\bar\xi_h)}{Through the readout state}\\
&\quad+\rterm{RelayTeal}{\frac{A_{hj}}{J}
(\varepsilon_j-\bar\varepsilon_h)}{Through explicit memory routing}.
\end{aligned}
\label{eq:relay-shared-gradient}
\end{equation}
\end{proposition}

The catalog cross-entropy residual and softmax Jacobian produce these two gradient paths (full derivation in Appendix~\ref{app:relay-gradient}): the readout term updates semantic matching and candidate gates, while the routing term compares each position's residual with its head average.

\subsection{Ranking Structure and One-Pass Recommendation}
\label{sec:relay-inference}

\paragraph{\textbf{Memory as a contextual probability tilt.}}
Let $r_\theta=f_\theta-b_\theta$ be the memory correction and let $p_0$ be the catalog softmax of $b_\theta$. Then
\begin{equation}
p_\theta(s\mid H;C)=
\frac{\rterm{RelayBlue}{p_0(s\mid H)}{Semantic prior}\,
\rterm{RelayTeal}{\exp r_\theta(s\mid H;C)}{Memory correction}}
{\rterm{RelayOrange}{\mathbb E_{v\sim p_0(\cdot\mid H)}
[\exp r_\theta(v\mid H;C)]}{Renormalization}}.
\label{eq:relay-tilt}
\end{equation}
For any $s,v\in\mathcal S$, memory changes their log odds by exactly $r_\theta(s)-r_\theta(v)$; it redistributes semantic preference at destinations supported by routed evidence.

\paragraph{\textbf{Sparse memory support.}}
The correction has support contained in
\begin{equation}
\begin{aligned}
\mathcal U_C(H)&=\bigcup_{j=1}^{m}
\left(\{s_j\}\cup\mathcal N_B^C(s_j)\right),\\
|\mathcal U_C(H)|&\leq m(B+1).
\end{aligned}
\label{eq:relay-support}
\end{equation}
Memory changes scores only within this support; candidates outside it retain their semantic scores. Recommendation ranks the corrected scores over the full catalog.

\paragraph{\textbf{One shared prediction path.}}
At serving time, RELAY encodes the history and evaluates its query once, scores the catalog with cached vectors, then adds routed memory evidence at its destinations. Semantic scoring costs $O(|\mathcal S|d)$; candidate gates add $O(m(B+1)d)$ on the sparse support, and the readout costs $O(md+d^2)$ when feed-forward width is proportional to $d$. The transition table stores $O(|\mathcal S|B)$ entries at inference and one additional entry per source during training. History length and memory budget control the additional scoring cost: behavioral evidence strengthens the ranking without a second catalog-wide neural evaluation. Algorithm~\ref{alg:relay} summarizes the procedure.

\begin{algorithm}[!t]
\caption{RELAY: shared-query learning and one-pass recommendation}
\label{alg:relay}
\small
\setlength{\tabcolsep}{2.5pt}
\renewcommand{\arraystretch}{1.04}
\begin{tabularx}{\linewidth}{@{}rX@{}}
\multicolumn{2}{@{}l@{}}{\textbf{Input:} records $\mathcal D_{\mathrm{tr}}$, catalog $\mathcal S$, budget $B$, epochs $T$, cutoff $K$.}\\
\multicolumn{2}{@{}l@{}}{\textbf{Output:} trained parameters $\theta$ and Top-$K$ recommendations.}\\[-1pt]
\rowcolor{RelayBlue!12}\multicolumn{2}{@{}l@{}}{\strut\textcolor{RelayBlue}{\textbf{Build transition memory}}}\\
1: & $C(a,b)\gets\sum_{n=1}^{N}\ind{s_{n,\mathrm{last}}=a,\,s_n^+=b,\,a\ne b}$ \hfill\textcolor{black!55}{\textit{// Eq.~(6)}}\\
2: & Store each source's top-$(B+1)$ positive neighbors and counts, with fixed tie order.\\
\rowcolor{RelayTeal!12}\multicolumn{2}{@{}l@{}}{\strut\textcolor{RelayTeal}{\textbf{Train the shared query and catalog scorer}}}\\
3: & \textbf{for} epoch $=1,\ldots,T$ \textbf{do}\\
4: & \hspace*{1em}\textbf{for} each mini-batch $\mathcal M\subset\mathcal D_{\mathrm{tr}}$ \textbf{do}\\
5: & \hspace*{2em}\textbf{for} each $(H_n,s_n^+)\in\mathcal M$ \textbf{do}\\
6: & \hspace*{3em}Compute query state $z_n$, shared attention $a_n$, and readout $u_\theta(H_n)$.\\
7: & \hspace*{3em}$C^{(-n)}\gets C$.\\
8: & \hspace*{3em}\textbf{if} $s_{n,\mathrm{last}}\ne s_n^+$ \textbf{then}\\
9: & \hspace*{4em}Decrement $C(s_{n,\mathrm{last}},s_n^+)$; recover its exact top-$B$ from the stored top-$(B+1)$.\\
10: & \hspace*{3em}\textbf{end if}\\
11: & \hspace*{3em}$\ell_n\gets-\log p_\theta(s_n^+\mid H_n;C^{(-n)})$ using the full-catalog softmax.\\
12: & \hspace*{2em}\textbf{end for}\\
13: & \hspace*{2em}$\theta\gets\operatorname{Update}\!\left(\theta,\frac{1}{|\mathcal M|}\sum_{n\in\mathcal M}\ell_n\right)$ \hfill\textcolor{black!55}{\textit{// Eq.~(13)}}\\
14: & \hspace*{1em}\textbf{end for}\\
15: & \textbf{end for}\\
\rowcolor{RelayOrange!12}\multicolumn{2}{@{}l@{}}{\strut\textcolor{RelayOrange}{\textbf{Recommend in one pass}}}\\
16: & Cache $\{v_\theta(s):s\in\mathcal S\}$ and the full-data top-$B$ transition memory.\\
17: & \textbf{for} each request history $H$ \textbf{do}\\
18: & \hspace*{1em}Compute $z$, $a$, and $u_\theta(H)$ once; set $f_\theta(s)\gets b_\theta(s\mid H)$ for all $s\in\mathcal S$.\\
19: & \hspace*{1em}\textbf{for} $s\in\mathcal U_C(H)$, add $\gamma_{\mathrm D}(s,z)m_{\mathrm D}(s\mid H)+\gamma_{\mathrm T}(s,z)m_{\mathrm T}(s\mid H;C)$ to $f_\theta(s)$.\\
20: & \hspace*{1em}\textbf{return} $\RelayTop_K\{f_\theta(s):s\in\mathcal S\}$.\\
21: & \textbf{end for}\\
\end{tabularx}
\end{algorithm}

\begin{table*}[t]
\centering
\caption{Full-catalog test Recall/NDCG@5. Baselines report means; RELAY reports mean $\pm$ standard deviation. Amber marks the strongest baseline and blue marks RELAY. Bold denotes the best mean; Improv.\ is relative to the strongest baseline.}
\label{tab:main-results}
\begingroup
\small
\renewcommand{\arraystretch}{1.10}
\OCMainTableFit{14}{ll*{12}{r}}{%
\toprule
& & \multicolumn{2}{c}{\textbf{Sports}} & \multicolumn{2}{c}{\textbf{Beauty}} & \multicolumn{2}{c}{\textbf{Toys}} & \multicolumn{2}{c}{\textbf{Scientific}} & \multicolumn{2}{c}{\textbf{Instruments}} & \multicolumn{2}{c}{\textbf{Games}} \\
\cmidrule(lr){3-4}\cmidrule(lr){5-6}\cmidrule(lr){7-8}\cmidrule(lr){9-10}\cmidrule(lr){11-12}\cmidrule(lr){13-14}
\rowcolor{OCTableHeader}
\textbf{Type} & \textbf{Method} & R@5 & N@5 & R@5 & N@5 & R@5 & N@5 & R@5 & N@5 & R@5 & N@5 & R@5 & N@5 \\
\midrule
\multirow{5}{*}{\textcolor{OCTableCategoryText}{\shortstack[l]{Item\\ID-based}}} & GRU4Rec~\cite{hidasi2016gru4rec}~(2015) & 0.0128 & 0.0087 & 0.0165 & 0.0098 & 0.0098 & 0.0060 & 0.0186 & 0.0127 & 0.0293 & 0.0195 & 0.0459 & 0.0310 \\
 & SASRec~\cite{kang2018sasrec}~(2018) & 0.0232 & 0.0155 & 0.0390 & 0.0248 & 0.0462 & 0.0304 & 0.0237 & 0.0150 & 0.0335 & 0.0212 & 0.0512 & 0.0326 \\
 & BERT4Rec~\cite{sun2019bert4rec}~(2019) & 0.0116 & 0.0074 & 0.0202 & 0.0125 & 0.0117 & 0.0070 & 0.0159 & 0.0101 & 0.0252 & 0.0158 & 0.0311 & 0.0198 \\
 & PreferDiff~\cite{liu2025preferdiff}~(2024) & 0.0276 & 0.0189 & 0.0453 & 0.0318 & 0.0605 & 0.0406 & 0.0163 & 0.0115 & 0.0210 & 0.0140 & 0.0353 & 0.0244 \\
 & FAVE~\cite{shi2026fave}~(2026) & 0.0303 & 0.0193 & 0.0583 & 0.0407 & 0.0615 & 0.0447 & 0.0229 & 0.0144 & 0.0324 & 0.0208 & 0.0544 & 0.0350 \\
\midrule
\multirow{5}{*}{\textcolor{OCTableCategoryText}{\shortstack[l]{SID-based\\(AR)}}} & TIGER~\cite{rajput2023tiger}~(2023) & 0.0267 & 0.0179 & 0.0455 & 0.0325 & 0.0519 & 0.0373 & 0.0279 & 0.0182 & 0.0360 & 0.0234 & 0.0528 & 0.0345 \\
 & ActionPiece~\cite{hou2025actionpiece}~(2025) & 0.0317 & 0.0203 & 0.0512 & 0.0341 & 0.0473 & 0.0311 & 0.0269 & 0.0172 & 0.0390 & 0.0245 & 0.0583 & 0.0381 \\
 & COBRA~\cite{yang2025cobra}~(2025) & 0.0301 & 0.0216 & 0.0539 & 0.0394 & 0.0621 & 0.0461 & 0.0275 & 0.0178 & 0.0345 & 0.0229 & 0.0473 & 0.0309 \\
 & SID-MLP~\cite{guo2026sidmlp}~(2026) & 0.0263 & 0.0174 & 0.0442 & 0.0303 & 0.0378 & 0.0243 & 0.0292 & 0.0194 & 0.0398 & 0.0254 & 0.0575 & 0.0381 \\
 & DIGER~\cite{fu2026diger}~(2026) & 0.0159 & 0.0101 & 0.0443 & 0.0286 & 0.0432 & 0.0291 & 0.0265 & 0.0167 & 0.0303 & 0.0190 & 0.0454 & 0.0285 \\
\midrule
\multirow{6}{*}{\textcolor{OCTableCategoryText}{\shortstack[l]{SID-based\\(NAR)}}} & RPG~\cite{hou2025rpg}~(2025) & 0.0312 & 0.0214 & 0.0546 & 0.0377 & 0.0593 & 0.0397 & 0.0259 & 0.0172 & 0.0358 & 0.0242 & 0.0582 & 0.0394 \\
 & LLaDA-Rec~\cite{shi2025lladarec}~(2025) & 0.0280 & 0.0186 & 0.0508 & 0.0348 & 0.0509 & 0.0345 & 0.0308 & 0.0202 & 0.0403 & 0.0265 & 0.0624 & 0.0414 \\
 & DiffGRM~\cite{liu2026diffgrm}~(2025) & 0.0366 & 0.0248 & 0.0605 & 0.0413 & 0.0617 & 0.0453 & 0.0404 & 0.0276 & 0.0430 & 0.0284 & 0.0619 & 0.0405 \\
 & OneGR~\cite{wang2026onegr}~(2026) & 0.0375 & 0.0254 & 0.0632 & 0.0430 & 0.0632 & 0.0462 & 0.0353 & 0.0244 & 0.0367 & 0.0248 & 0.0545 & 0.0366 \\
 & SPRINT~\cite{cai2026sprint}~(2026) & 0.0413 & 0.0271 & 0.0674 & 0.0452 & 0.0705 & 0.0500 & 0.0442 & 0.0300 & 0.0460 & 0.0306 & 0.0635 & 0.0426 \\
\rowcolor{OCTableBestBaseline}
 & OneChord~\cite{li2026onechord}~(2026) & 0.0438 & 0.0295 & 0.0731 & 0.0497 & 0.0746 & 0.0546 & 0.0475 & 0.0325 & 0.0480 & 0.0319 & 0.0671 & 0.0435 \\
\midrule
\rowcolor{OCTableOurs}
\textcolor{OCTableCategoryText}{Ours} & \textbf{RELAY} & \OCMeanSD{0.05087}{0.0007} & \OCMeanSD{0.03530}{0.0006} & \OCMeanSD{0.08121}{0.0014} & \OCMeanSD{0.05854}{0.0010} & \OCMeanSD{0.08897}{0.0016} & \OCMeanSD{0.06576}{0.0014} & \OCMeanSD{0.05005}{0.0010} & \OCMeanSD{0.03507}{0.0006} & \OCMeanSD{0.05448}{0.0006} & \OCMeanSD{0.03673}{0.0004} & \OCMeanSD{0.08049}{0.0007} & \OCMeanSD{0.05495}{0.0005} \\
\addlinespace[1pt]
& \textit{Improv.} & 16.14\% & 19.66\% & 11.09\% & 17.79\% & 19.26\% & 20.44\% & 5.37\% & 7.91\% & 13.50\% & 15.14\% & 19.96\% & 26.32\% \\
\bottomrule
}
\endgroup

\end{table*}

\begin{table*}[t]
\centering
\caption{Full-test inference cost (s) and RELAY's speedup over each baseline. Amber marks the lowest baseline cost; blue marks RELAY. Bold denotes the lowest cost.}
\label{tab:inference-cost}
\begingroup
\small
\renewcommand{\arraystretch}{1.19}
\OCMainTableFit{13}{l*{12}{r}}{%
\toprule
 & \multicolumn{2}{c}{\textbf{Sports}} & \multicolumn{2}{c}{\textbf{Beauty}} & \multicolumn{2}{c}{\textbf{Toys}} & \multicolumn{2}{c}{\textbf{Scientific}} & \multicolumn{2}{c}{\textbf{Instruments}} & \multicolumn{2}{c}{\textbf{Games}} \\
\cmidrule(lr){2-3}\cmidrule(lr){4-5}\cmidrule(lr){6-7}\cmidrule(lr){8-9}\cmidrule(lr){10-11}\cmidrule(lr){12-13}
\rowcolor{OCTableHeader}
\textbf{Method} & Cost & Speedup & Cost & Speedup & Cost & Speedup & Cost & Speedup & Cost & Speedup & Cost & Speedup \\
\midrule
\rowcolor{OCTableBestBaseline}
FAVE~\cite{shi2026fave}~(2026) & 0.3437 & 1.36$\times$ & 0.2098 & 1.38$\times$ & 0.1818 & 1.38$\times$ & 0.5110 & 1.36$\times$ & 0.5705 & 1.32$\times$ & 0.9486 & 1.35$\times$ \\
TIGER~\cite{rajput2023tiger}~(2023) & 10.0199 & 39.67$\times$ & 6.2677 & 41.08$\times$ & 5.4396 & 41.19$\times$ & 14.3327 & 38.16$\times$ & 16.1564 & 37.45$\times$ & 26.7062 & 37.97$\times$ \\
ActionPiece~\cite{hou2025actionpiece}~(2025) & 39.2035 & 155.20$\times$ & 24.7523 & 162.25$\times$ & 21.4593 & 162.51$\times$ & 56.9978 & 151.73$\times$ & 63.8786 & 148.07$\times$ & 105.9439 & 150.64$\times$ \\
RPG~\cite{hou2025rpg}~(2025) & 0.7240 & 2.87$\times$ & 0.5641 & 3.70$\times$ & 0.3615 & 2.74$\times$ & 1.6441 & 4.38$\times$ & 1.8188 & 4.22$\times$ & 3.0426 & 4.33$\times$ \\
LLaDA-Rec~\cite{shi2025lladarec}~(2025) & 85.1093 & 336.93$\times$ & 56.8194 & 372.45$\times$ & 45.6964 & 346.05$\times$ & 121.7687 & 324.16$\times$ & 136.7625 & 317.00$\times$ & 228.0063 & 324.21$\times$ \\
DiffGRM~\cite{liu2026diffgrm}~(2025) & 2.1590 & 8.55$\times$ & 1.3601 & 8.92$\times$ & 1.1404 & 8.64$\times$ & 3.0965 & 8.24$\times$ & 3.4338 & 7.96$\times$ & 5.7371 & 8.16$\times$ \\
SPRINT~\cite{cai2026sprint}~(2026) & 0.5007 & 1.98$\times$ & 0.2844 & 1.86$\times$ & 0.2477 & 1.88$\times$ & 0.7681 & 2.04$\times$ & 0.8861 & 2.05$\times$ & 1.4658 & 2.08$\times$ \\
OneChord~\cite{li2026onechord}~(2026) & 0.4423 & 1.75$\times$ & 0.2732 & 1.79$\times$ & 0.2337 & 1.77$\times$ & 0.6700 & 1.78$\times$ & 0.7629 & 1.77$\times$ & 1.2441 & 1.77$\times$ \\
\midrule
\rowcolor{OCTableOurs}
\textbf{RELAY} & \textbf{0.2526} & --- & \textbf{0.1526} & --- & \textbf{0.1320} & --- & \textbf{0.3756} & --- & \textbf{0.4314} & --- & \textbf{0.7033} & --- \\
\bottomrule
}
\endgroup

\end{table*}

\begin{table*}[t]
\centering
\caption{Full-catalog test results for RELAY ablations and model performance under three quantization methods. The lower panel compares SPRINT, OneChord, and RELAY on Sports, Beauty, and Toys; means weight these datasets equally. The nonnegative controls are single-run references.}
\label{tab:ablation}
\begingroup
\small
\renewcommand{\arraystretch}{1.12}
\OCMainTableFit{13}{l*{12}{r}}{%
\toprule
 & \multicolumn{2}{c}{\textbf{Sports}} & \multicolumn{2}{c}{\textbf{Beauty}} & \multicolumn{2}{c}{\textbf{Toys}} & \multicolumn{2}{c}{\textbf{Scientific}} & \multicolumn{2}{c}{\textbf{Instruments}} & \multicolumn{2}{c}{\textbf{Games}} \\
\cmidrule(lr){2-3}\cmidrule(lr){4-5}\cmidrule(lr){6-7}\cmidrule(lr){8-9}\cmidrule(lr){10-11}\cmidrule(lr){12-13}
\rowcolor{OCTableHeader}
\textbf{Variant} & R@5 & N@5 & R@5 & N@5 & R@5 & N@5 & R@5 & N@5 & R@5 & N@5 & R@5 & N@5 \\
\midrule
\rowcolor{OCTableOurs}
\textbf{RELAY} & 0.05087 & 0.03530 & 0.08121 & 0.05854 & 0.08897 & 0.06576 & 0.05005 & 0.03507 & 0.05448 & 0.03673 & 0.08049 & 0.05495 \\
\midrule
w/o behavioral memory & 0.04465 & 0.02955 & 0.07430 & 0.05151 & 0.07980 & 0.05690 & 0.04625 & 0.03183 & 0.05045 & 0.03282 & 0.07307 & 0.04842 \\
w/o transition memory & 0.04567 & 0.02985 & 0.07357 & 0.05141 & 0.07738 & 0.05559 & 0.04515 & 0.03162 & 0.04962 & 0.03251 & 0.07210 & 0.04766 \\
w/o direct-history memory & 0.04985 & 0.03439 & 0.08007 & 0.05739 & 0.08736 & 0.06425 & 0.04885 & 0.03426 & 0.05355 & 0.03591 & 0.07941 & 0.05386 \\
w/o current-record exclusion & 0.03558 & 0.02511 & 0.06821 & 0.05099 & 0.07658 & 0.05861 & 0.03143 & 0.02352 & 0.03515 & 0.02459 & 0.06543 & 0.04592 \\
Separate memory routing & 0.04995 & 0.03439 & 0.08038 & 0.05770 & 0.08776 & 0.06485 & 0.04825 & 0.03416 & 0.05345 & 0.03580 & 0.07931 & 0.05375 \\
\midrule
Original nonnegative control & 0.05085 & 0.03533 & 0.08040 & 0.05706 & 0.08783 & 0.06516 & 0.04949 & 0.03459 & 0.05254 & 0.03556 & 0.07531 & 0.05081 \\
Candidate-conditioned nonnegative control & 0.05096 & 0.03540 & 0.08161 & 0.05752 & 0.08773 & 0.06551 & 0.04933 & 0.03468 & 0.05315 & 0.03587 & 0.07537 & 0.05120 \\
\bottomrule
}
\endgroup\par
\vspace{3pt}
\begingroup
\small
\renewcommand{\arraystretch}{1.06}
\OCMainTableFit{10}{ll*{8}{r}}{%
\toprule
 & & \multicolumn{2}{c}{\textbf{Sports}} & \multicolumn{2}{c}{\textbf{Beauty}} & \multicolumn{2}{c}{\textbf{Toys}} & \multicolumn{2}{c}{\textbf{Mean}} \\
\cmidrule(lr){3-4}\cmidrule(lr){5-6}\cmidrule(lr){7-8}\cmidrule(lr){9-10}
\rowcolor{OCTableHeader}
\textbf{Method} & \textbf{Model} & R@5 & N@5 & R@5 & N@5 & R@5 & N@5 & R@5 & N@5 \\
\midrule
\multirow{3}{*}{OPQ} & SPRINT & 0.04110 & 0.02700 & 0.06750 & 0.04510 & 0.07080 & 0.05010 & 0.05980 & 0.04073 \\
 & OneChord & 0.04380 & 0.02950 & 0.07310 & 0.04970 & 0.07460 & 0.05460 & 0.06383 & 0.04460 \\
\rowcolor{OCTableOurs}
\cellcolor{white} & \textbf{RELAY} & \textbf{0.05087} & \textbf{0.03530} & \textbf{0.08121} & \textbf{0.05854} & \textbf{0.08897} & \textbf{0.06576} & \textbf{0.07368} & \textbf{0.05320} \\
\midrule
\multirow{3}{*}{RQ-VAE} & SPRINT & 0.02840 & 0.01860 & 0.05050 & 0.03480 & 0.04520 & 0.03070 & 0.04137 & 0.02803 \\
 & OneChord & 0.02871 & 0.01825 & 0.04932 & 0.03319 & 0.04863 & 0.03136 & 0.04222 & 0.02760 \\
\rowcolor{OCTableOurs}
\cellcolor{white} & \textbf{RELAY} & \textbf{0.03840} & \textbf{0.02696} & \textbf{0.06484} & \textbf{0.04639} & \textbf{0.07166} & \textbf{0.05176} & \textbf{0.05830} & \textbf{0.04171} \\
\midrule
\multirow{3}{*}{RQ-Kmeans} & SPRINT & 0.02960 & 0.01950 & 0.05180 & 0.03520 & 0.05460 & 0.03550 & 0.04533 & 0.03007 \\
 & OneChord & 0.02809 & 0.01804 & 0.05017 & 0.03374 & 0.04832 & 0.03186 & 0.04219 & 0.02788 \\
\rowcolor{OCTableOurs}
\cellcolor{white} & \textbf{RELAY} & \textbf{0.03806} & \textbf{0.02652} & \textbf{0.06573} & \textbf{0.04696} & \textbf{0.07212} & \textbf{0.05278} & \textbf{0.05864} & \textbf{0.04209} \\
\bottomrule
}
\endgroup\par
\end{table*}

\begin{table}[t]
\centering
\caption{Predictor parameters across datasets and RELAY's parameter reduction. Amber marks the lowest baseline mean.}
\label{tab:parameters}
\begingroup
\small
\renewcommand{\arraystretch}{1.08}
\OCMainTableFit{3}{lrr}{%
\toprule
\rowcolor{OCTableHeader}
\textbf{Method} & \textbf{Parameters (M)} & \textbf{Reduction} \\
\midrule
FAVE & 2.998--4.780 & 31.84--57.25\% \\
TIGER & 4.591 & 55.49\% \\
ActionPiece & 9.580 & 78.67\% \\
RPG & 8.515--13.568 & 76.00--84.94\% \\
LLaDA-Rec & 4.749 & 56.97\% \\
\rowcolor{OCTableBestBaseline}
DiffGRM & 3.237 & 36.86\% \\
SPRINT & 6.184 & 66.96\% \\
OneChord & 5.860 & 65.12\% \\
\midrule
\rowcolor{OCTableOurs}
\textbf{RELAY} & \textbf{2.044} & -- \\
\bottomrule
}
\endgroup
\end{table}

\begin{table}[t]
\caption{Full-catalog test performance by interaction sparsity, averaged equally over six datasets. Gains are relative to OneChord.}
\label{tab:sparsity}
\centering
\begingroup
\small
\renewcommand{\arraystretch}{1.12}
\OCMainTableFit{7}{l*{6}{r}}{%
\toprule
\rowcolor{OCTableHeader}
 & \multicolumn{2}{c}{\textbf{OneChord}} & \multicolumn{2}{c}{\textbf{RELAY}} & \multicolumn{2}{c}{Gain} \\
\cmidrule(lr){2-3}\cmidrule(lr){4-5}\cmidrule(lr){6-7}
\rowcolor{OCTableHeader}
\textbf{Group} & R@5 & N@5 & R@5 & N@5 & R@5 & N@5 \\
\midrule
\multicolumn{7}{l}{\textit{Target-item popularity}} \\
Tail & 0.0276 & 0.0198 & \textbf{0.0293} & \textbf{0.0208} & +6.06\% & +4.92\% \\
Middle & 0.0408 & 0.0293 & \textbf{0.0467} & \textbf{0.0342} & +14.44\% & +16.66\% \\
Head & 0.0851 & 0.0569 & \textbf{0.0993} & \textbf{0.0695} & +16.67\% & +22.04\% \\
\midrule
\multicolumn{7}{l}{\textit{Effective history length}} \\
Short & 0.0616 & 0.0419 & \textbf{0.0705} & \textbf{0.0499} & +14.48\% & +18.96\% \\
Medium & 0.0570 & 0.0389 & \textbf{0.0647} & \textbf{0.0453} & +13.50\% & +16.68\% \\
Long & 0.0545 & 0.0377 & \textbf{0.0646} & \textbf{0.0455} & +18.64\% & +20.60\% \\
\bottomrule
}
\endgroup
\end{table}

\begin{figure*}[t]
\centering
\includegraphics[width=\textwidth]{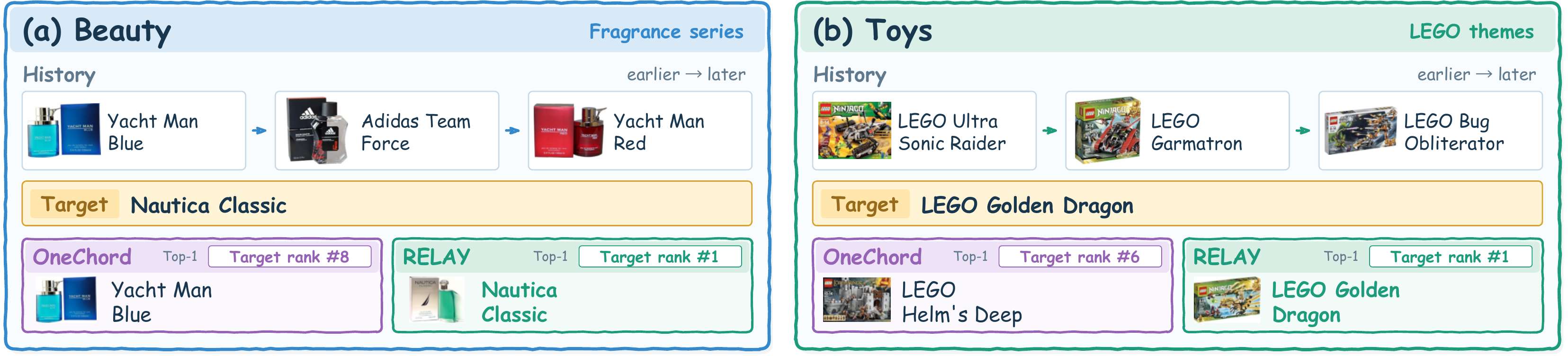}\par
\includegraphics[width=.88\textwidth]{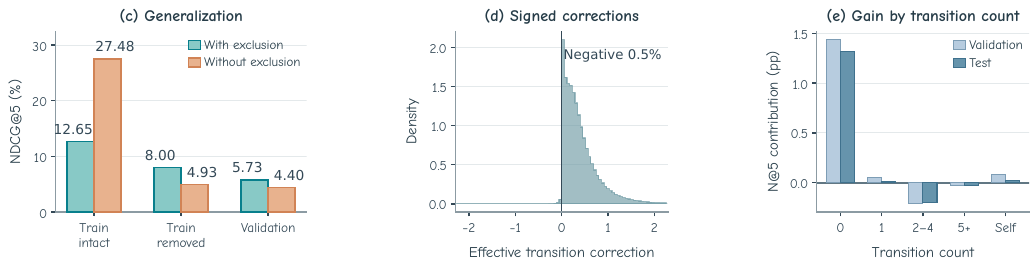}
\caption{Recommendation examples (a,b) and exclusion diagnostics (c--e).}
\Description{The upper panel shows two recommendation examples. RELAY ranks the Beauty target Nautica Classic first while OneChord ranks it eighth, and RELAY ranks the Toys target LEGO The Golden Dragon first while OneChord ranks it sixth. The lower-left panel compares full-catalog N@5 for models trained with and without current-record exclusion on intact and record-removed training windows and validation windows. The lower-middle panel plots the distribution of effective signed transition corrections and marks zero. The lower-right panel compares validation and test N@5 gain contributions for transition counts zero, one, two to four, five or more, and self transitions.}
\label{fig:exclusion-mechanism}
\end{figure*}

\section{Experiments}
\label{sec:experiments}

We evaluate RELAY's ranking accuracy, predictor size, and inference cost. Ablations and routing interventions test shared history selection, while record-exclusion diagnostics examine generalization beyond recorded transitions.

\subsection{Experimental Setup}
\label{sec:experimental-setup}

\paragraph{Datasets and evaluation.}
We use Sports, Beauty, and Toys from Amazon Reviews 2014~\cite{mcauley2015amazon}, and Scientific, Instruments, and Games from Amazon Reviews 2023~\cite{hou2026amazon}. We report Recall@5 (R@5) and NDCG@5 (N@5) under full-catalog evaluation. For SID-based ranking, the catalog consists of all distinct legal four-code identifiers, and a hit requires an exact match to the held-out target SID. Inference cost measures generation of the final Top-10 lists over the complete test set. Dataset statistics, chronological splits, and evaluation definitions appear in Appendix~\ref{app:datasets}.

\paragraph{Baselines.}
We compare RELAY with $16$ methods spanning item ID-based recommendation, autoregressive SID-based generation, and non-autoregressive SID-based prediction (Table~\ref{tab:main-results}). OneChord provides the strongest accuracy baseline. The inference-cost comparison covers eight baselines (Table~\ref{tab:inference-cost}).

\paragraph{Implementation.}
RELAY uses a fixed semantic catalog, a bidirectional history encoder, and a single query readout. The shared history attention serves both semantic matching and behavioral-memory routing, trained through the complete-catalog likelihood in Eq.~\eqref{eq:relay-objective}. Architecture, optimization, random seeds, checkpoint selection, and measurement settings are specified in Appendix~\ref{app:experimental-protocol}.

\subsection{Overall Recommendation Performance}
\label{sec:overall-performance}

\paragraph{Consistent gains across domains.}
Table~\ref{tab:main-results} shows that RELAY achieves the highest mean R@5 and N@5 on every dataset. Relative to OneChord, the average improvements are \textbf{14.22\%} and \textbf{17.88\%}, respectively. N@5 gains range from $7.91\%$ on Scientific to $26.32\%$ on Games.

\paragraph{Higher-ranked relevant predictions.}
RELAY improves both target retrieval and its position near the top. On Toys, R@5 increases from $0.0746$ to $0.0890$ and N@5 from $0.0546$ to $0.0658$. Compared with SPRINT, RELAY improves R@5 and N@5 by $21.38\%$ and $26.20\%$ on average.

\subsection{Inference Efficiency and Model Size}
\label{sec:inference-efficiency}

\paragraph{One forward pass with faster inference.}
RELAY completes full-test Top-10 generation in \textbf{0.13--0.70 seconds} (Table~\ref{tab:inference-cost}), with a \textbf{1.32--1.38$\times$} speedup over one-step FAVE, the fastest compared baseline. Its speedup is $1.75$--$1.79\times$ over OneChord and $1.86$--$2.08\times$ over SPRINT. It combines the highest mean accuracy with the lowest inference cost among the compared models.

\paragraph{A compact recommendation network.}
RELAY has \textbf{2.044 million parameters}, the smallest predictor in Table~\ref{tab:parameters}. It reduces parameters by \textbf{65.12\%} relative to OneChord, \textbf{66.96\%} relative to SPRINT, and $31.84$--$57.25\%$ relative to FAVE. Shared routing retains semantic and behavioral evidence within a compact network. Appendix~\ref{app:efficiency-protocol} gives the protocol; Appendix~\ref{app:multi-query} evaluates a multi-query extension.

\subsection{Ablation Study}
\label{sec:ablation}

Table~\ref{tab:ablation} evaluates the two memory channels, current-record exclusion, and attention sharing. Its lower panel compares SPRINT, OneChord, and RELAY with three quantizers on Sports, Beauty, and Toys: RELAY leads both metrics in every setting. Data and optimization are matched; Appendix~\ref{app:ablation-protocol} defines each intervention.

\paragraph{Two memory channels strengthen semantic ranking.}
Removing all behavioral memory reduces average R@5 and N@5 by $9.25\%$ and $12.33\%$. Removing transition memory while retaining direct-history memory reduces them by $10.49\%$ and $13.17\%$. Removing direct-history memory reduces them by $1.72\%$ and $2.20\%$. Both channels contribute in every domain, with transition memory providing the larger gain.

\paragraph{Excluding a record's own contribution improves generalization.}
Removing current-record exclusion reduces average R@5 by \textbf{23.07\%} and N@5 by \textbf{20.12\%}, the largest ablation effect. Separate routing reduces them by $1.72\%$ and $1.99\%$.

\paragraph{Shared attention improves accuracy with fewer parameters.}
Sharing saves $65{,}556$ parameters by reusing the semantic readout's selector, while retaining higher average accuracy in all six domains. The same interaction weights determine both the semantic preference and the behavioral associations supplied to each candidate. The ranking objective therefore trains one history selection through both roles, directly coupling preference representation with the evidence used to rank candidates.

\subsection{Performance across Interaction Sparsity}
\label{sec:sparsity-analysis}

RELAY's advantage extends across target popularity and the amount of available history. Table~\ref{tab:sparsity} compares RELAY with OneChord~\cite{li2026onechord} under full-catalog test ranking, averaging each group's metrics equally over the six datasets. The two partitions distinguish evidence available for a target from context available for a user. Groups use target-item training frequency and effective history length; Appendix~\ref{app:sparsity-analysis} specifies the partitions.

\paragraph{Gains span the popularity spectrum.}
RELAY improves N@5 by \textbf{4.92\%}, \textbf{16.66\%}, and \textbf{22.04\%} for tail, middle, and head targets. R@5 improves by $6.06\%$, $14.44\%$, and $16.67\%$. Gains remain positive across the popularity groups.

\paragraph{Gains persist across history lengths.}
For short, medium, and long histories, N@5 improves by \textbf{18.96\%}, \textbf{16.68\%}, and \textbf{20.60\%}; the corresponding R@5 gains are $14.48\%$, $13.50\%$, and $18.64\%$. Short histories test evidence recovery from limited context, while long histories test whether relevant interactions remain identifiable among more alternatives. The gains in both groups show that shared history selection supports accurate ranking across different amounts of available context.

\subsection{Query-Guided Memory Routing}
\label{sec:query-routing-analysis}

\begin{table}[b]
\caption{History routing interventions with fixed predictor parameters. Full-catalog test metrics average six domains equally.}
\label{tab:query-routing}
\centering
\begingroup
\small
\renewcommand{\arraystretch}{1.12}
\OCMainTableFit{4}{lrrr}{%
\toprule
\rowcolor{OCTableHeader}
\textbf{Routing} & R@5 & N@5 & $\Delta$N@5 \\
\midrule
\rowcolor{OCTableOurs}
\textbf{Learned query} & \textbf{0.06768} & \textbf{0.04773} & --- \\
Uniform & 0.06364 & 0.04404 & -7.73\% \\
Recency & 0.06700 & 0.04711 & -1.29\% \\
Shuffled & 0.06146 & 0.04239 & -11.18\% \\
\bottomrule
}
\endgroup
\end{table}

\textbf{The query's attention selects useful behavioral evidence.} On trained RELAY predictors, replacing query-guided memory routing with uniform weights reduces N@5 by \textbf{7.73\%}; shuffling the learned weights across valid history positions reduces it by \textbf{11.18\%} (Table~\ref{tab:query-routing}). The semantic readout and candidate-conditioned gates remain fixed. Shuffling preserves the weight values while changing their historical sources, identifying the correspondence between an interaction and its routing weight as a source of ranking accuracy.

\textbf{Content-dependent selection becomes more valuable in longer histories.} Relative to position-based recency routing, the query's N@5 gains increase from $0.58\%$ for short histories to $1.96\%$ for medium histories and \textbf{2.84\%} for long histories. As the set of historical interactions grows, the learned selection supplies increasingly useful evidence for candidate ranking. Appendix~\ref{app:query-routing-analysis} defines the routing interventions and aggregation.

\subsection{Record Exclusion and Test Cases}
\label{sec:exclusion-mechanism}

Figure~\ref{fig:exclusion-mechanism} shows the effect of exact record exclusion: without it, training N@5 reaches $27.48\%$ (vs. $12.65\%$) as validation falls to $4.40\%$ (vs. $5.73\%$); removing each target transition reverses the training ranking ($4.93\%$ vs. $8.00\%$). Signed updates include negative evidence ($0.53\%$), while unseen transitions account for most held-out gain. Memory complements semantic scores with evidence routed from relevant history interactions beyond the last item. In the examples, RELAY ranks Nautica Classic and LEGO Golden Dragon first, while OneChord ranks them eighth and sixth. Appendix~\ref{app:case-examples} gives Top-3 recommendations and examples from four more domains.

\section{Conclusion}
\label{sec:conclusion}

RELAY realizes \textbf{one query, two roles}: the same history attention forms semantic preferences and routes behavioral evidence into candidate scores. One catalog likelihood trains both roles, with candidate-conditioned gates calibrating direct-history and transition evidence. Across six domains, RELAY improves Recall@5 and NDCG@5 over the strongest baseline by $14.22\%$ and $17.88\%$ on average, with $65.12\%$ fewer predictor parameters and $1.75$--$1.79\times$ faster inference. Shared history selection thus unifies preference representation and behavioral evidence within a compact recommendation network.

\bibliographystyle{ACM-Reference-Format}
\providecommand{\showeprint}[2][]{\href{https://arxiv.org/abs/#2}{arXiv:#2}}
\bibliography{references}

\appendix
\section{Derivations and Additional Method Details}
\label{app:relay-method}

\subsection{Representation and Prediction Space}
\label{app:relay-notation}
\label{app:relay-computation}

All scores, counts, likelihoods, and rankings use the distinct SID catalog $\mathcal S$. History sums include valid positions only; padding has zero attention. The $L$ digits occupy separate ranges of one embedding table. History codes, candidate codes, and $L$ copies of the readout state use the shared projection
\begin{equation}
M_\theta(x)=W_2\operatorname{ReLU}(W_1x+b_1)+b_2,
\label{eq:relay-shared-mlp}
\end{equation}
where $W_1\in\mathbb R^{d\times Ld}$ and $W_2\in\mathbb R^{d\times d}$. Candidate code embeddings stop gradients; both uses train the projection. The encoder is bidirectional and pre-normalized. The readout projects a normalized query, takes encoded history head slices as keys and values, and applies an output projection, residual feed-forward block, and final RMS normalization. Routing averages pre-dropout head probabilities. Candidate vectors are cached for fixed parameters; memory stores $B$ destinations per source for serving and one additional destination for training. The two candidate-conditioned gates use $2d+2$ parameters, the same count as the contextual strengths.

\subsection{Exact Current-Record Exclusion}
\label{app:relay-exclusion}

\paragraph{Proof of Proposition~\ref{prop:relay-exclusion}.}
Order a source's positive-count destinations as $v_1,v_2,\ldots$ by decreasing count and a fixed tie order. Decrementing $v_r$ preserves every other pairwise order. If $r>B$, the top-$B$ is unchanged. If $r\leq B$, at most that entry leaves; among the remaining destinations, $v_{B+1}$ precedes every possible replacement. The updated top-$B$ therefore lies within the original top-$(B+1)$. Re-sort these entries with adjusted counts and the same tie order, discard zero counts, and return at most $B$ entries. This recovers the exact updated neighborhood.\hfill$\square$

For record $n$, the off-diagonal entry $(s_{n,\mathrm{last}},s_n^+)$ loses one count. Every occurrence of the same source in $H_n$ reads this adjusted row. Self transitions contribute no count. Exclusion removes one record's contribution; other training windows, including the same user's windows, remain. Counts use training records alone. Serving uses the original top-$B$ table.

\subsection{Gradient of Shared Attention}
\label{app:relay-gradient}

\paragraph{Proof of Proposition~\ref{prop:relay-shared-gradient}.}
Fix the encoded history and $C^{(-n)}$. Treat readout state $z$ and routing weights $a$ as separate arguments:
\begin{equation}
\begin{aligned}
F_s(z,a)&=b_s(z)+\sum_j a_jR_j(s;z),\\
R_j(s;z)&=\gamma_{\mathrm D}(s,z)d_j(s)
+\gamma_{\mathrm T}(s,z)t_j(s;C^{(-n)}).
\end{aligned}
\label{eq:relay-gradient-score}
\end{equation}
Set $\delta_s=p_s-\ind{s=s^+}=\partial\ell/\partial F_s$. The state derivative and its attention contribution are
\begin{equation}
\begin{aligned}
g_z&=\sum_{s\in\mathcal S}\delta_s
\left[\nabla_zb_s+\sum_j a_j\nabla_zR_j(s;z)\right],\\
\xi_{hj}&=\left\langle g_z,\partial z/\partial A_{hj}\right\rangle.
\end{aligned}
\label{eq:relay-state-gradient}
\end{equation}
Routing contributes $\varepsilon_j=\sum_s\delta_sR_j(s;z)$. Since $a_j=J^{-1}\sum_h A_{hj}$,
\begin{equation}
\frac{\partial\ell}{\partial A_{hj}}=\xi_{hj}+\frac{\varepsilon_j}{J}.
\label{eq:relay-attention-gradient}
\end{equation}
Applying $\partial A_{hk}/\partial\eta_{hj}=A_{hk}(\ind{k=j}-A_{hj})$ to Eq.~\eqref{eq:relay-attention-gradient} gives Eq.~\eqref{eq:relay-shared-gradient}. The identity also holds conditional on fixed dropout; query and encoder gradients include their other paths.\hfill$\square$

The residual $\varepsilon_j=\mathbb E_{s\sim p_\theta}[R_j(s;z)]-R_j(s^+;z)$ compares target evidence with its model-weighted average. Shared attention receives this routing signal together with the state derivative through the same logits.

\subsection{Probability Tilt}

Writing $Z_0=\sum_s e^{b_\theta(s)}$ and $Z=\sum_s e^{b_\theta(s)+r_\theta(s)}$ gives $Z/Z_0=\mathbb E_{p_0}[e^{r_\theta(s)}]$, establishing Eq.~\eqref{eq:relay-tilt}. Hence
\begin{equation}
\log\frac{p_\theta(s)}{p_\theta(v)}
-\log\frac{p_0(s)}{p_0(v)}=r_\theta(s)-r_\theta(v).
\label{eq:relay-odds-shift}
\end{equation}
Candidates outside the correction support preserve their pairwise odds and order; normalization remains catalog-wide. The implementation scores the full catalog, then adds the sparse memory corrections before selecting the final Top-$K$.

\section{Experimental Protocol}
\label{app:experimental-protocol}

\begin{figure}[t]
\centering
\includegraphics[width=\linewidth]{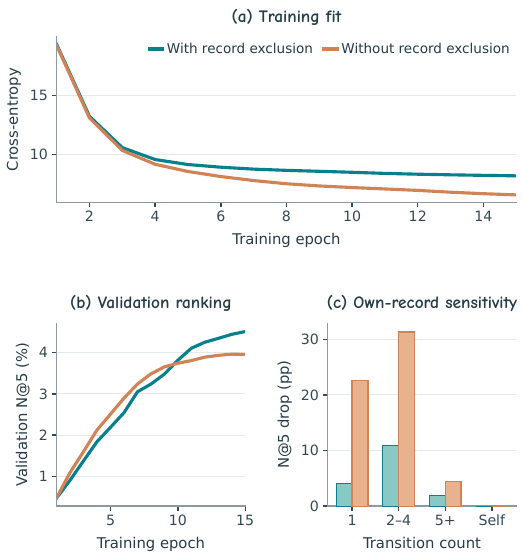}
\caption{Training and own-record sensitivity for candidate-conditioned signed RELAY. (a) Six-domain mean catalog cross-entropy during training. (b) Mean full-catalog validation N@5 across common epochs. (c) Training N@5 drop after removing each record's own transition, grouped by transition count.}
\Description{Panel a plots equal-domain-averaged training cross-entropy for models trained with and without current-record exclusion. Panel b plots their equal-domain-averaged full-catalog validation N@5 over the shared training epochs. Panel c shows the equal-domain-averaged N@5 difference between intact and record-removed training graphs at each selected checkpoint, grouped by transition counts one, two to four, five or more, and self.}
\label{fig:exclusion-self-evidence}
\end{figure}

\subsection{Datasets and Evaluation}
\label{app:datasets}

\begin{table}[t]
\caption{Dataset statistics after preprocessing. SIDs are distinct four-code identifiers; Avg.\ is mean sequence length.}
\label{tab:dataset-statistics}
\centering
\small
\setlength{\tabcolsep}{2.8pt}
\begin{tabular}{@{}lrrrrr@{}}
\toprule
Dataset & Users & Items & SIDs & Interactions & Avg.\\
\midrule
Sports & 35,598 & 18,357 & 15,102 & 296,337 & 8.32\\
Beauty & 22,363 & 12,101 & 10,377 & 198,502 & 8.88\\
Toys & 19,412 & 11,924 & 10,220 & 167,597 & 8.63\\
Scientific & 50,985 & 25,848 & 19,810 & 412,947 & 8.10\\
Instruments & 57,439 & 24,587 & 20,951 & 511,836 & 8.91\\
Games & 94,762 & 25,612 & 20,944 & 814,586 & 8.60\\
\bottomrule
\end{tabular}
\end{table}

Amazon 2014 uses category-specific 5-core data. For Amazon 2023, we retain each user's longest chronologically ordered interaction sequence. The last two interactions are validation/test targets; windows use the earlier prefix. Histories retain at most $20$ preceding interactions.

All methods are evaluated over the complete set of distinct four-code SIDs using exact matches to the held-out target SID. For item-ID baselines, we map the full-catalog item ranking through the fixed OPQ mapping and collapse repeated SIDs at their first-ranked position before computing the metrics. Items with the same SID occupy one candidate entry. For target rank $r_u$ among $U$ users,
\begin{equation}
\begin{aligned}
\mathrm{R@}K&=U^{-1}\sum_u\ind{r_u\leq K},\\
\mathrm{N@}K&=U^{-1}\sum_u\frac{\ind{r_u\leq K}}{\log_2(r_u+1)}.
\end{aligned}
\label{eq:relay-eval-metrics}
\end{equation}
Table~\ref{tab:relay-k10} supplements the main $K=5$ comparison.

\begin{table}[t]
\centering
\caption{RELAY full-catalog test performance at $K=10$.}
\label{tab:relay-k10}
\small
\begin{tabular}{lcc}
\toprule
Dataset & R@10 & N@10 \\
\midrule
Sports & 0.07245 & 0.04223 \\
Beauty & 0.11139 & 0.06829 \\
Toys & 0.11720 & 0.07491 \\
Scientific & 0.07279 & 0.04238 \\
Instruments & 0.07920 & 0.04470 \\
Games & 0.11755 & 0.06686 \\
\bottomrule
\end{tabular}
\end{table}

\paragraph{Item-level recommendation.}
Table~\ref{tab:item-level-results} compares the strongest item-ID, autoregressive SID, and non-autoregressive SID baselines, using full-catalog item-level metrics averaged over Sports, Beauty, and Toys. RELAY improves mean R@5/N@5 by $12.88\%$/$15.32\%$ over the strongest baseline.

\begin{table}[t]
\caption{Item-level full-catalog results averaged over Sports, Beauty, and Toys. Each baseline leads its recommendation paradigm.}
\label{tab:item-level-results}
\centering
\small
\renewcommand{\arraystretch}{1.08}
\OCMainTableFit{4}{llrr}{%
\toprule
\rowcolor{OCTableHeader}
Paradigm & Method & R@5 & N@5\\
\midrule
\rowcolor{OCTableBestBaseline}
Item-ID & FAVE~\cite{shi2026fave} & 0.04883 & 0.03408\\
SID-based (AR) & ActionPiece~\cite{hou2025actionpiece} & 0.03530 & 0.02200\\
SID-based (NAR) & OneGR~\cite{wang2026onegr} & 0.04490 & 0.03210\\
\midrule
\rowcolor{OCTableOurs}
Ours & \textbf{RELAY} & \textbf{0.05512} & \textbf{0.03930}\\
\bottomrule
}
\end{table}

\subsection{Architecture and Training}
\label{app:implementation}

\paragraph{Predictor and memory.}
Each SID has four $256$-value digits. RELAY uses width $256$, four heads, feed-forward width $1024$, dropout $0.1$, one bidirectional encoder layer, and one cross-attention layer without query self-attention. The shared projection outputs $256$ dimensions; $\tau=0.01$, and candidate embeddings stop gradients. Memory retains $B=8$ off-diagonal destinations with $\log(1+C)$ weights and a ninth exclusion backup. A $256$-to-$2$ affine gate (514 parameters) conditions memory corrections on the candidate; mean pre-dropout attention routes both channels.

\paragraph{Optimization.}
Training uses fresh initialization, AdamW ($10^{-4}$ learning rate, zero weight decay), gradient clipping at $1$, batch size $1024$, and cosine scheduling after $10{,}000$ warmup steps. Catalog cross-entropy trains for at most $150$ epochs with per-epoch validation and patience $10$; evaluation batch size is $256$.

\subsection{Seeds and Aggregation}
\label{app:baseline-protocol}
\label{app:optimization-timing}

Main results average five seeds; all experiments other than the main and ablation tables use seed $2024$. Checkpoints maximize validation $0.8\,\mathrm{N@10}+0.2\,\mathrm{R@10}$ before full-test evaluation. RELAY reports mean and sample standard deviation (denominator $4$); baselines report means. Main improvements average six domain-relative gains between RELAY and each baseline:
\begin{equation}
G_m=\frac{100}{6}\sum_d
\left(\frac{\bar m_{d,\mathrm{RELAY}}}{\bar m_{d,\mathrm{baseline}}}-1\right).
\label{eq:relay-relative-gain}
\end{equation}
This gives $14.22\%$ R@5 and $17.88\%$ N@5 over OneChord. Grouped gains compare equally averaged group metrics.

\subsection{Efficiency and Parameters}
\label{app:efficiency-protocol}

Timing uses NVIDIA A800 80GB PCIe, FP32 without TF32, batch size $256$, GPU-resident inputs, and cached candidate vectors. Each model uses its official inference configuration and is measured with six warmed-up full-test CUDA Graph passes; we report median latency. Timed passes include the complete inference path through Top-10 selection. Model loading, host transfer, data preparation, and warmup are excluded. All other measurement settings follow the RELAY protocol. Each model uses its checkpoint trained on the corresponding dataset. Speedup divides baseline time by RELAY time.

We count predictor scalars and embeddings once, excluding tokenizers, training-only components, memory, and caches. RELAY has $2{,}043{,}650$ parameters ($514$ for strengths); RPG has $8{,}514{,}944$ on Sports/Toys and $13{,}568{,}384$ elsewhere. Parameter reductions are relative to each baseline. Figure~\ref{fig:relay-teaser} averages domain N@5, parameters, and inference seconds, with logarithmic cost.

\subsection{Ablation Definitions}
\label{app:ablation-protocol}

Table~\ref{tab:ablation} reports full-catalog results. Memory/channel deletions remove their corresponding corrections. Without exclusion trains with $C$ rather than $C^{(-n)}$; both serve with $C$. Separate routing adds a $256\times256$ projection and $20$ positional biases: $2{,}109{,}206$ parameters, a $65{,}556$ ($3.11\%$) difference. The quantizer panel changes only the four-code assignments, using the same Sentence-T5/PCA item vectors and training settings for OPQ, RQ-VAE, and residual K-means. The nonnegative rows are single-run controls.

\begin{figure*}[t]
\centering
\includegraphics[width=\textwidth]{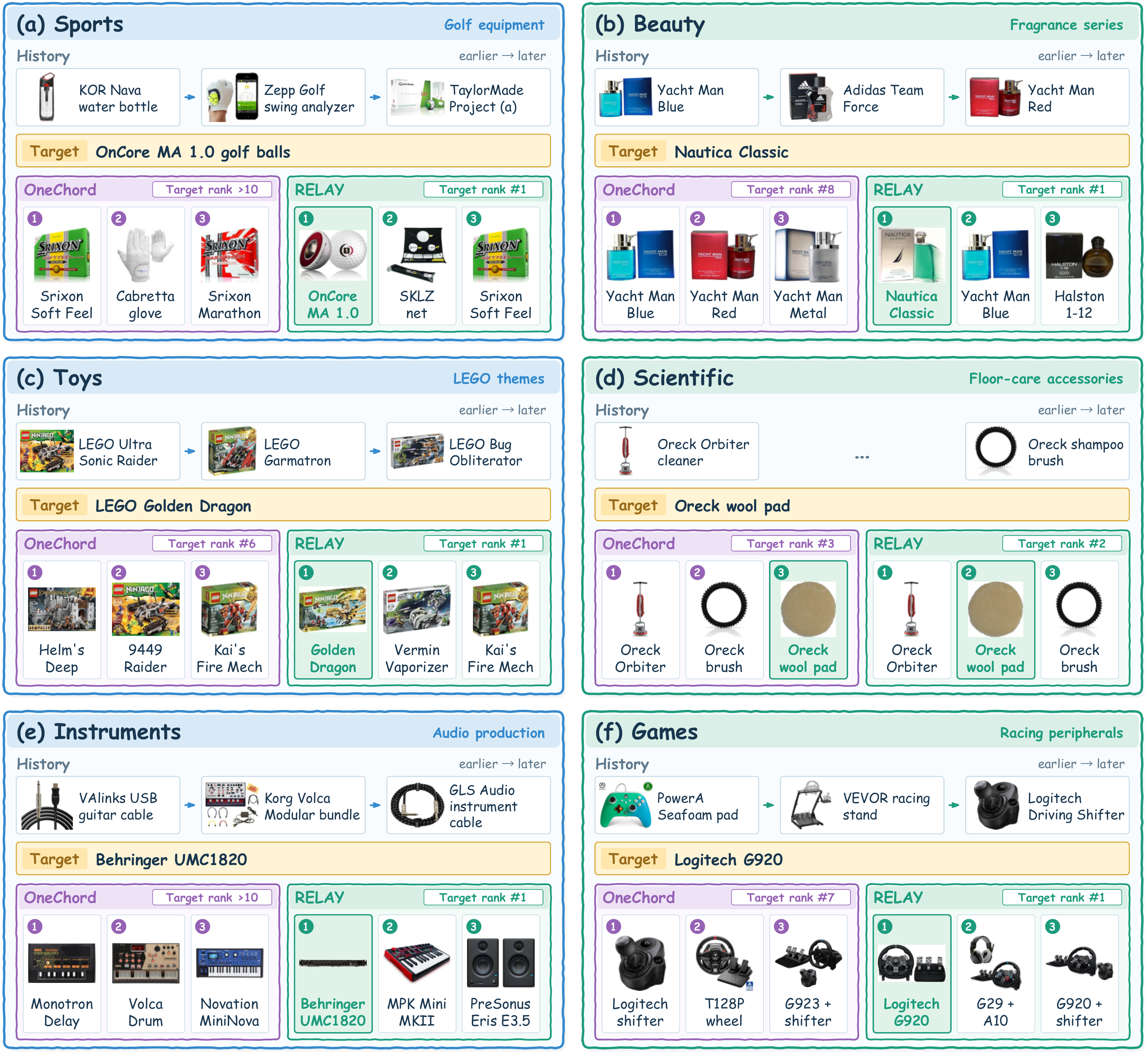}
\caption{Full-catalog recommendation cases across six domains.}
\Description{Six product-photo examples compare OneChord and RELAY. Target ranks improve from beyond ten to first on Sports, eighth to first on Beauty, sixth to first on Toys, third to second on Scientific, beyond ten to first on Instruments, and seventh to first on Games.}
\label{fig:relay-cases-appendix}
\end{figure*}

\subsection{Multi-Query Extension}
\label{app:multi-query}

With $Q$ learned queries, RELAY shares the encoder, cross-attention, projection, and strength predictor. Each query performs semantic readout and memory routing; the maximum complete query score ranks each candidate. Table~\ref{tab:multi-query} averages six-domain full-test results. Two/four queries improve N@5 by $4.10\%$/$5.08\%$ and long-history N@5 by $3.85\%$/$4.94\%$, adding $256$/$768$ parameters and $11.28\%$/$33.93\%$ inference time.

\begin{table}[t]
\caption{Six-domain means for the multi-query extension. Long histories contain $11$--$20$ interactions; inference cost follows Appendix~\ref{app:efficiency-protocol}.}
\label{tab:multi-query}
\centering
\small
\OCMainTableFit{6}{rrrrrr}{%
\toprule
\rowcolor{OCTableHeader}
$Q$ & R@5 & N@5 & Long N@5 & Cost (s) & Params (M)\\
\midrule
\rowcolor{OCTableOurs}
1 & 0.06768 & 0.04773 & 0.04549 & \textbf{0.3413} & \textbf{2.04365}\\
2 & 0.07010 & 0.04968 & 0.04724 & 0.3797 & 2.04391\\
4 & \textbf{0.07076} & \textbf{0.05015} & \textbf{0.04774} & 0.4571 & 2.04442\\
\bottomrule
}
\end{table}

\subsection{Interaction Sparsity}
\label{app:sparsity-analysis}

Table~\ref{tab:sparsity} pairs validation-selected models. Raw-item frequency counts each training-prefix interaction once; upward-rounded catalog percentiles preserve ties: tail $f\leq q_{50}$, middle $q_{50}<f\leq q_{80}$, head $f>q_{80}$. Thresholds are $(7,15)$ Sports/Beauty, $(6,14)$ Toys, $(7,14)$ Scientific, $(8,17)$ Instruments, and $(9,25)$ Games. Short/medium/long histories contain $\leq5$/$6$--$10$/$11$--$20$ interactions. Both partitions cover $280{,}559$ test users; exact-SID metrics average domains equally, and gains use unrounded means.

\subsection{Query Routing Intervention}
\label{app:query-routing-analysis}

Table~\ref{tab:query-routing} changes only memory attention: uniform $1/m$, recency proportional to $1/(1+t-j)$ for final position $t$, or five shuffles of learned weights. Padding stays zero; shuffling preserves weights and normalization. Full-catalog evaluation covers all $280{,}559$ users with equal-domain aggregation and the groups in Section~\ref{app:sparsity-analysis}.

\subsection{Record-exclusion Diagnostics}
\label{app:exclusion-diagnostics}

Figure~\ref{fig:exclusion-self-evidence} compares matched models over $52$ shared epochs. The final panel removes each training record's own transition at the selected checkpoint, grouping N@5 changes by that transition's remaining frequency.

\section{Illustrated Recommendation Cases}
\label{app:case-examples}

Figure~\ref{fig:relay-cases-appendix} shows full-catalog Top-3 recommendations with histories of up to $20$ interactions. Each displayed item has a unique SID. RELAY ranks the target first in five domains and second in Scientific; OneChord ranks it $>10$, $8$, $6$, $3$, $>10$, and $7$ in Sports, Beauty, Toys, Scientific, Instruments, and Games. Histories are shortened for display.

In Beauty, OneChord's Top-3 stays within Yacht Man fragrances, while RELAY ranks Nautica Classic first. In Toys, RELAY places Golden Dragon ahead of other LEGO sets, distinguishing the exact continuation within the same theme.

Equipment histories lead to golf balls, an audio interface, and a racing wheel. In Scientific, RELAY places the target wool pad ahead of the brush.

\end{document}